\documentclass[prd,amsfonts,aps,nofootinbib,notitlepage,10pt,superscriptaddress]{revtex4-1}
\pdfoutput=1
\usepackage{amsmath,amssymb}
\usepackage{graphicx}
\usepackage[utf8]{inputenc}
\usepackage{cancel}
\usepackage[colorlinks=true]{hyperref}
\usepackage{color}
\usepackage{array,multirow} 
\usepackage{subcaption}

\renewcommand{\arraystretch}{1.1}

\newcommand{\UdeAaddress}{Instituto de F\'isica, Universidad de Antioquia, Calle 70 No. 52-21, Medell\'in, Colombia.}

\begin{document}
\title{One-loop Dirac neutrino mass and mixed axion/WIMP dark matter}
\author{Cristian D. R. Carvajal}\email{cdavid.ruiz@udea.edu.co}
\author{Óscar Zapata}\email{oalberto.zapata@udea.edu.co}
\affiliation{$ $~\UdeAaddress}
\date{\today}
          
\begin{abstract}
  We consider the Peccei-Quinn (PQ) mechanism as the one behind the Dirac neutrino masses when these are generated through the $d=5$ effective operator $\bar{L}\tilde{H}N_R\phi$ at one loop level, with $\phi$ being a Standard Model singlet scalar.
  In this setup, the PQ symmetry guarantees that the one-loop realization of such an effective operator gives the leading contribution to the Dirac neutrino masses by forbidding the contributions arising from its tree level realizations. 
  All the mediators in the one-loop neutrino mass diagrams can be stabilized by a remnant $ Z_N$ symmetry from the PQ symmetry breaking, thus forming a dark sector besides the axion sector and leading to mixed axion/WIMP dark matter scenarios. 
\end{abstract}

\maketitle

\section{Introduction}
In the last decades the neutrino oscillations have been firmly established thanks to a great dedicated experimental program \cite{Tanabashi:2018oca}, leading to a concise and a clear understanding of the  neutrino oscillation pattern \cite{deSalas:2017kay}.
However, this wisdom has not shed light on the underlying mechanism behind the neutrino masses and the properties of neutrinos under the particle-antiparticle conjugation operation.
This adds to the negative results regarding the Majorana nature of the neutrinos from neutrinoless double-beta decay experiments \cite{KamLAND-Zen:2016pfg,Agostini:2018tnm,Aalseth:2017btx,Alduino:2017ehq,Albert:2017owj,Arnold:2016bed}.
Hence, a growing interest in Dirac neutrinos mass models has recently appeared, especially in models that contain a weakly interacting massive particle (WIMP) \cite{Steigman:1984ac} as the dark matter candidate that is not trivially connected with the neutrino mass generation mechanism \cite{Farzan:2012sa,Ma:2014qra,Ma:2015raa,Ma:2015mjd,Ma:2016mwh,Wang:2016lve,Bonilla:2016zef,Chulia:2016ngi,Borah:2016zbd,Wang:2017mcy,Yao:2017vtm,Yao:2018ekp,Ma:2017kgb,Wang:2017mcy,Reig:2018mdk,Chulia:2016ngi,CentellesChulia:2018gwr}.

It worth noting here  that the main motivation for considering WIMPs is  that they lie at the scale which at the physics beyond the Standard Model (SM) is expected to appear, thus leading to signals in dedicated experiments looking for dark matter (DM).  However, the WIMP paradigm is not free of theoretical and experimental challenges \cite{Baer:2014eja}, such as the lack of signals in those experiments so far, which have led to severe constraints over the parameter space of WIMP models \cite{Bertone:2010at,Baer:2014eja,Escudero:2016gzx,Arcadi:2017kky}.
Therefore, it would be relevant to consider approaches beyond the WIMP paradigm such as the multi-component DM \cite{Zurek:2008qg,Profumo:2009tb}, where the DM of the Universe is composed by, for instance,  WIMPs and QCD axions \cite{Baer:2011hx,Bae:2013hma,Dasgupta:2013cwa,Alves:2016bib,Ma:2017zyb,Chatterjee:2018mac}. 

Within the framework of the SM it is not possible to form a Dirac mass term for the neutrinos because of the lack of the right-handed partners of  active neutrinos $\nu_L$. Once they are added, Dirac neutrino masses are generated via the $d=4$ operator
\begin{equation}\label{eq:opdim4}
 \mathcal{O}_4= y\,\overline{L}\widetilde{H}N_{R}+\textrm{h.c.,}
\end{equation}
where $N_{R}$ are right-handed neutrinos, $L=(\nu_L,\ell_L)^T$ is the left-handed lepton doublet and $H=(H^+,H^0)^T$ is the Higgs doublet. 
Under the lepton number conservation (to protect the Diracness of neutrinos) this operator leads to sub-eV neutrino masses for $|y|\lesssim10^{-13}$, thus leaving unsettled the explanation of the smallness of the neutrino mass scale. 
From this, it would be reasonable to forbid $\mathcal{O}_4$ through a certain symmetry $\mathcal{X}$  while generating Dirac neutrino masses via higher dimensional operators,  either at  tree  level  or  loop level. 
On those lines, if the Higgs doublet is the only scalar in the particle spectrum the lowest dimensional operator would be of dimension $d=6$, while if a scalar SM singlet $\phi$  charged under $\mathcal{X}$ is added then a $d=5$ operator $\mathcal{O}_5$ would arise \cite{Ma:2016mwh,Yao:2018ekp,CentellesChulia:2018gwr,CentellesChulia:2018bkz}.    
\begin{equation}\label{eq:opdim5}
   \mathcal{O}_5=y'\overline{L}\widetilde{H}N_{R}\phi+\textrm{h.c.}.
\end{equation}
It follows that after the spontaneous breaking of $\mathcal{X}$ (due to the vacuum expectation value of $\phi$) an effective $\mathcal{O}_4$-like operator will arise but with the benefit that $\mathcal{O}_4$ itself is forbidden. 
Additionally,  the $\mathcal{X}$ symmetry may also may serve as the stabilizing symmetry of possible WIMP candidates arising from the one-loop (ultraviolet) realizations of $\mathcal{O}_5$ \cite{Ma:2015mjd,Ma:2016mwh,Wang:2016lve,Bonilla:2016zef,Chulia:2016ngi,Borah:2016zbd,Wang:2017mcy}. 

In this work, we associate the Peccei-Quinn (PQ) mechanism \cite{Peccei:1977hh} to scotogenic models with one-loop Dirac neutrino masses\footnote{See Refs.~\cite{Mohapatra:1982tc,Shafi:1984ek,Langacker:1986rj,Shin:1987xc,He:1988dm,Berezhiani:1989fp,Bertolini:1990vz,Ma:2001ac,Chen:2012baa,Dasgupta:2013cwa,Bertolini:2014aia,Gu:2016hxh,Ma:2017zyb,Ma:2017vdv,Suematsu:2017kcu,Suematsu:2017hki,Reig:2018ocz,Reig:2018yfd} for scenarios where the PQ mechanism is deeply related to the neutrino mass generation (at tree or loop level).}  through identification  of $\phi$ in $\mathcal{O}_5$ as the scalar field hosting the QCD axion and the PQ symmetry as responsible for the absence of $\mathcal{O}_4$ and the stability of the possible WIMP candidates.
It follows that after the spontaneous electroweak and PQ symmetry breaking  Dirac neutrino masses will be generated via the $\mathcal{O}_5$ operator, and imposing that $H$ does not carry a PQ charge (while $L$ and/or $N_R$ do) the contribution to the neutrino masses from the $\mathcal{O}_4$ operator is automatically forbidden, and under some PQ charge assignments it is also possible to make the one loop realization of $\mathcal{O}_5$ the main source of neutrino masses.  That is, the contributions of the three Dirac seesaw mechanisms are no longer present and the smallness of neutrino masses is related to the radiative character besides the large mass suppression coming from the loop mediators.
Furthermore, we will show that the same PQ charge assignments lead to a residual $Z_N$ discrete symmetry after the PQ symmetry breaking that guarantees the stability of the lightest of the mediators in the one-loop neutrino mass diagrams, thus leading naturally to  axion/WIMP DM scenarios. 

\section{Framework}\label{sec:framework}
The apparent non-observation of CP violation originating from the $\theta$ term in the QCD Lagrangian is a strong theoretical motivation for going beyond the SM, since it can be dynamically explained through the PQ mechanism \cite{Peccei:1977hh}.
Such a mechanism  requires one to extend the SM gauge group with an anomalous global symmetry, $U(1)_\textrm{PQ}$, which is spontaneously broken at some high scale \cite{Kim:1979if,Shifman:1979if,Dine:1981rt,Zhitnitsky:1980tq} by the vacuum expectation value $v_S$ of the scalar $S$ that hosts the QCD axion $a(x)$ \cite{Weinberg:1977ma,Wilczek:1977pj},
\begin{align}
&  S=\frac{1}{\sqrt{2}}[\rho(x)+v_S]e^{ia(x)/v_S}.
\end{align}
Here $\rho(x)$ is the radial part that will gain a mass of order of the  PQ symmetry breaking scale, which is constrained by several astrophysical phenomena such as supernova cooling \cite{Raffelt:2006cw} and black hole superradiance \cite{Arvanitaki:2014wva} to   $v_S\sim[10^9,10^{17}]$ GeV.  As canonical axion model we consider a hadronic KSVZ type model \cite{Kim:1979if,Shifman:1979if}, that is, we add to the SM the axionic field $S$ and two chiral singlet quarks $D_L$ and $D_R$, all of them charged under the PQ symmetry and interacting through the Yukawa term  $y_D S \overline{D_L}D_R$.

In the present framework $S$ additionally plays the role of $\phi$ in the $\mathcal{O}_5$ operator, with both $L$ and $N_R$ charged under the PQ symmetry whereas $H$ does not in such a way $\mathcal{O}_5$ is allowed and $\mathcal{O}_4$ is banned.
Contrary to the original KSVZ model, where all the SM fermions are neutral under the PQ symmetry, in this framework the leptons do have PQ charges but the quarks remain neutral. The  lepton number L conservation  is further imposed with the usual assignment for the SM fields (1 for the leptons and 0 for the rest), 1 for $N_R$ and 0 for $S$,  in order to prevent the appearance of the Majorana mass term $N_RN_R$ and the $d=5$ Weinberg operator $LHLH$, along their induced partners $N_RN_RS$ and  $LHLHS$, and thus  protecting the Dirac nature of neutrinos.
However, it should be noted that by allowing large values for the PQ charges and not imposing L conservation is possible to obtain a consistent model where the Diracness of neutrinos is guaranteed~\cite{Chen:2012baa} (see Sec.~\ref{sec. particular case} for a specific example). 

More specifically, the $S$, $H$ and $N_R$ fields transform under the PQ symmetry as
\begin{align}\label{PQ transformation}
  & S\rightarrow e^{i\mathcal{X}_S\xi}S, \ \ \ \ \ H\rightarrow H, \ \ \ \ \ N_R\rightarrow e^{i\mathcal{X}_{N_R}\xi}N_R,\hspace{1cm}\textrm{with}\,\,\,\mathcal{X}_S\neq0. 
\end{align}
It follows that the PQ charge of the lepton doublet must be $\mathcal{X}_L=\mathcal{X}_{N_R}+\mathcal{X}_S$ in order to simultaneously make the $\mathcal{O}_5$ operator PQ-invariant and  $\mathcal{O}_4$ forbidden.  Regarding the PQ charges of $D_R$ and $D_L$, they are related through the Yukawa term $y_D S\overline{D_L}D_R$ so  $\mathcal{X}_{D_L}=\mathcal{X}_{D_R}+\mathcal{X}_S$. This, in turn, implies that the color anomaly coefficient $\mathcal{C}$ of the resulting axion models is set by $\mathcal{C}=|\mathcal{X}_S|/2=M$, with $M$ a positive integer\footnote{The color anomaly is given by $\mathcal{C}=\sum_i|\mathcal{X}_{D_L}-\mathcal{X}_{D_R}|T(\mathcal{C}_i)$, where the sum is over all the irreducible $SU(3)_C\times U(1)_\textrm{EM}$ representations and $T(\mathcal{C}_i)$ is the index of the $SU(3)_C$ representation of the $D_{L,R}$ fields  ($T(3)=1/2$) \cite{Srednicki:1985xd}. Note that the condition of $\mathcal{C}$ to be an integer is mandatory to guarantee the periodicity of the axion potential, $2\pi f_a=2\pi v_S/\mathcal{C}$, and the interpretation of the axion as the phase of the field $S$ (implying a periodicity of $2\pi v_S$) \cite{Marsh:2015xka,Raffelt:1990yz}.
  Notice here that $v_S$ is related to  the axion decay constant $f_a$ via $f_a= v_S/\mathcal{C}$, where the color anomaly also sets the number of vacua of the $S$ (the so called domain wall number \cite{Marsh:2015xka}).}, {\it i.e.} the PQ charge of $S$ must be even. 
On the other hand, we assume $\mathcal{X}_{D_R}\neq0$ to avoid the mixing term $\overline{q_L}HD_R$ (the SM quark sector is not charged under the $U(1)_\textrm{PQ}$). All in all, these L and PQ charge assignments automatically forbid $\mathcal{O}_4$ and the Majorana mass terms for $N_R$ and $\nu_L$, and leave invariant $\mathcal{O}_5$.
However, they do not necessarily prevent the contributions to the neutrino mass matrix arising from  the tree-level realizations of $\mathcal{O}_5$.

\begin{figure}[t]
\begin{subfigure}[b]{0.25\textwidth}
        \includegraphics[scale=0.5]{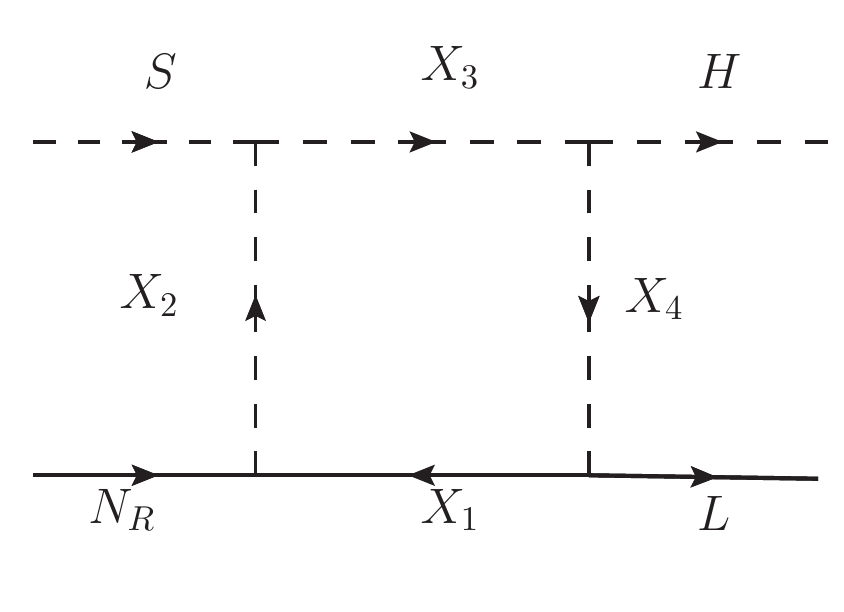}
        \caption{T1-1-A model}\label{fig:T11A}
  \end{subfigure}
  \begin{subfigure}[b]{0.25\textwidth}
        \includegraphics[scale=0.5]{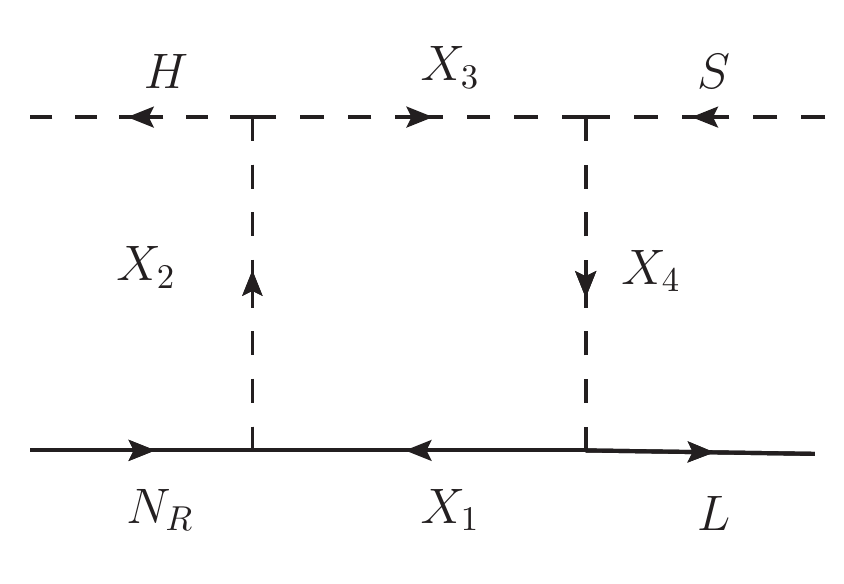}
        \caption{T1-1-B model}\label{fig:T11B}
  \end{subfigure}
  \begin{subfigure}[b]{0.25\textwidth}
        \includegraphics[scale=0.5]{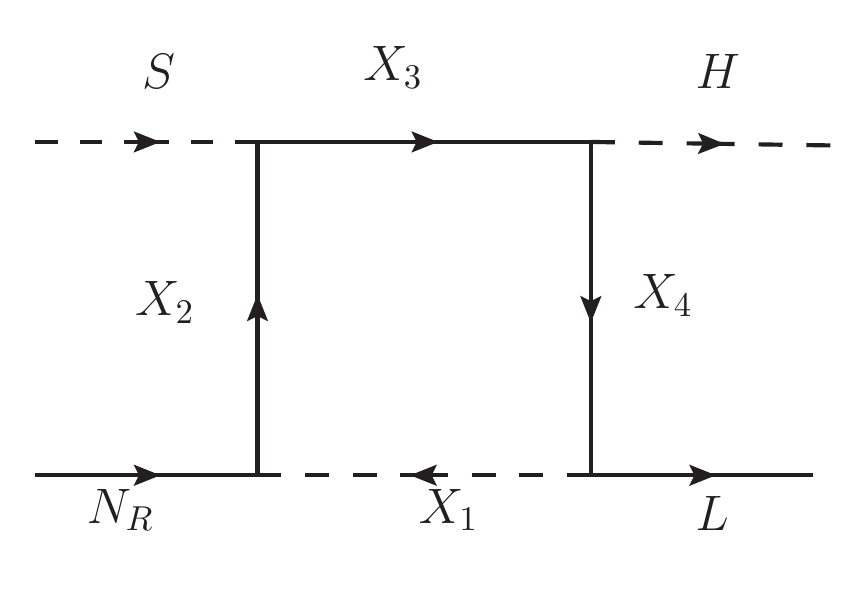}
        \caption{T1-2-A model}\label{fig:T12A}
  \end{subfigure}
  \begin{subfigure}[b]{0.25\textwidth}
        \includegraphics[scale=0.5]{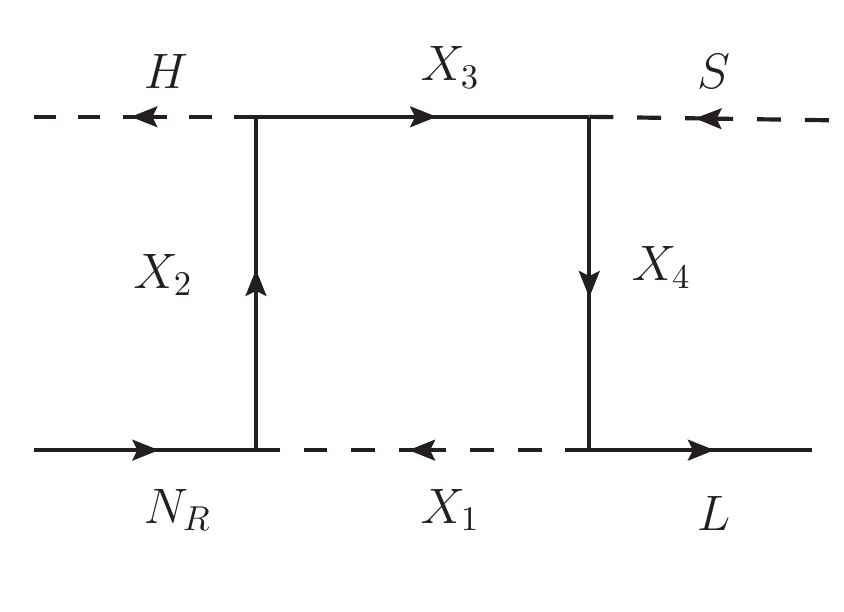}
        \caption{T1-2-B model}\label{fig:T12B}
  \end{subfigure}
  \begin{subfigure}[b]{0.25\textwidth}
        \includegraphics[scale=0.5]{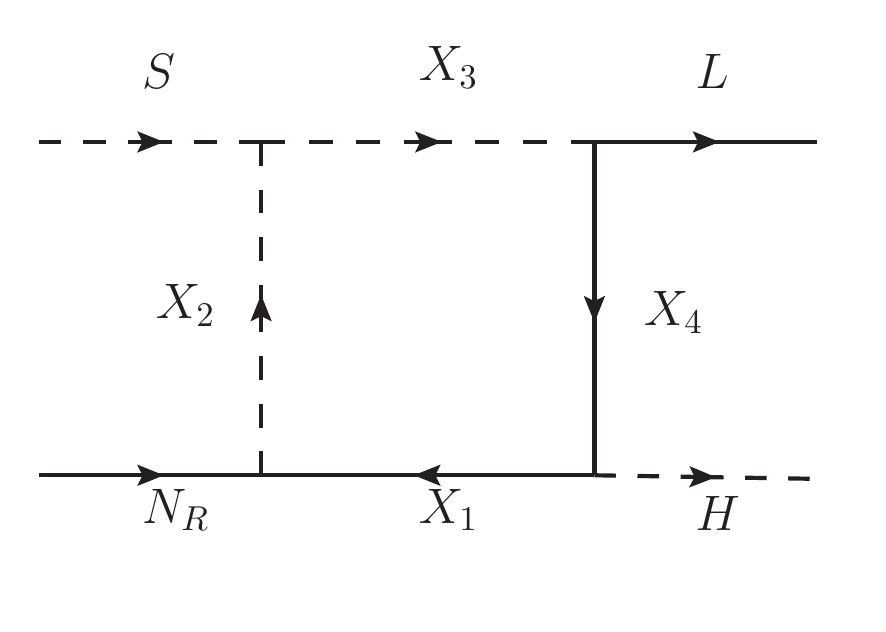}
        \caption{T1-3-D model}\label{fig:T13D}
  \end{subfigure}
  \begin{subfigure}[b]{0.25\textwidth}
        \includegraphics[scale=0.5]{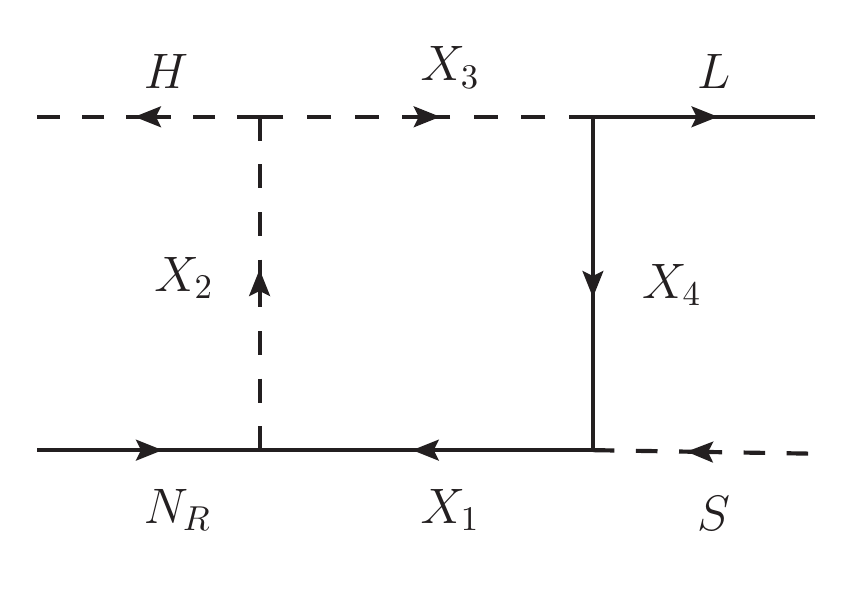}
        \caption{T1-3-E model}\label{fig:T13E}
      \end{subfigure}
  \begin{subfigure}[b]{0.25\textwidth}
        \includegraphics[scale=0.5]{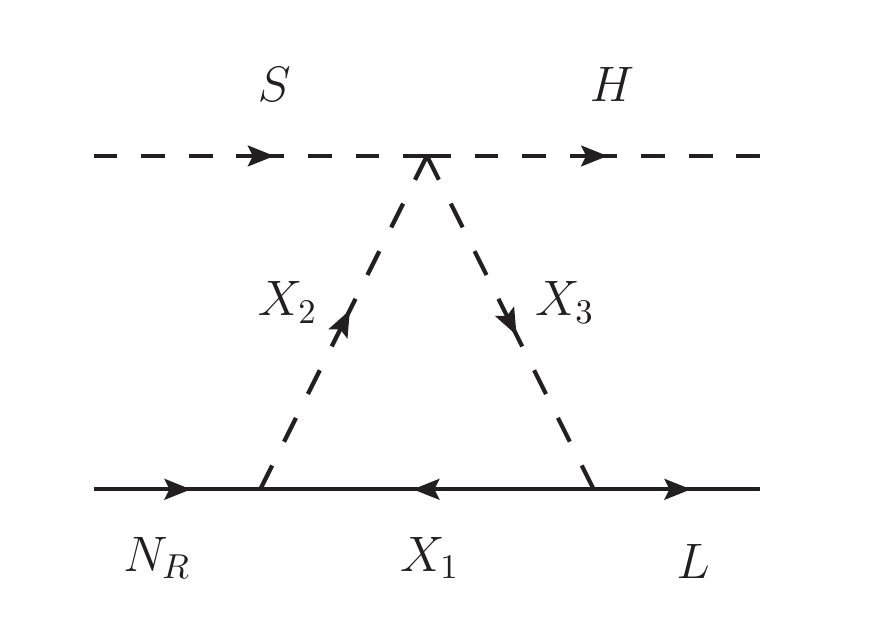}
        \caption{T3-1-A model}\label{fig:T13E}
      \end{subfigure}
  \begin{subfigure}[b]{0.25\textwidth}
  \includegraphics[scale=0.5]{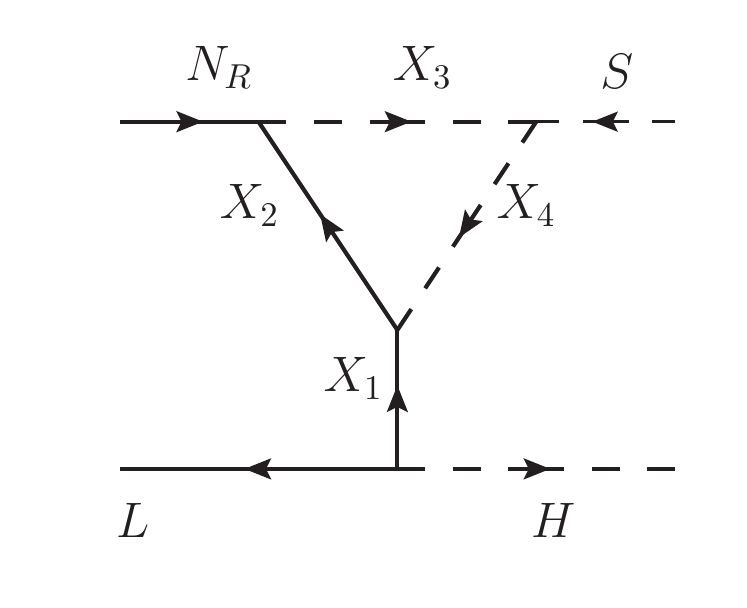}
        \caption{T4-3-I model}\label{fig:T13E}
      \end{subfigure}
\caption{One-loop diagrams with T1, T3 and T4 topologies leading to Dirac neutrino masses. }
\label{fig:T1-topology}
\end{figure}

In the presence of right-handed neutrinos and a scalar singlet, the tree-level realizations  of $\mathcal{O}_5$ demand the introduction of either a SM singlet vector-like fermion, an extra Higgs doublet or a vector-like lepton doublet as mediator fields, which lead to the three Dirac seesaw mechanisms \cite{Ma:2016mwh,Yao:2018ekp,CentellesChulia:2018gwr,CentellesChulia:2018bkz}: type I, type II and type III, respectively.
On the other hand,  all possible realizations -considering new fields (fermions and scalars) transforming as singlets, doublets or triplets under $SU(2)_{L}$, and color singlets,
with arbitrary values of the hypercharge  but fixed up to a free parameter $\alpha$\footnote{The hypercharge of the mediators is set to $Y=\alpha$ or $Y=\alpha\pm1$ so that the possible values for $\alpha$ are 0,1,2 in order to have $SU(2)_L$ multiplets featuring  a neutral particle.}- of $\mathcal{O}_5$ leading to Dirac neutrino masses at one-loop level were studied in Ref.~\cite{Yao:2018ekp}. There it was obtained that a finite list of sets of mediator fields exists, which give rise to neutrino masses at one-loop level, classified according to the topology of the diagram. 
In that analysis, a $Z_2$ symmetry was used to forbid $\mathcal{O}_4$, with both $N_R$ and $S$ odd and the SM sector even under this $Z_2$, and the new fermions were assumed to be vector-like to ensure anomaly cancellation of the SM gauge symmetries.
We will use their results regarding the renormalizable and genuine one-loop models to study the necessary conditions to implement the PQ mechanism as the responsible  behind the one-loop Dirac neutrino masses\footnote{We will closely follow the notation introduced in Ref.~\cite{Yao:2018ekp} for the topologies, fields, and genuine models.}.  

At one-loop level, there are three different topologies that lead to renormalizable and genuine one-loop realizations~\cite{Yao:2018ekp}: the T1 (box-like), T3 (triangle-like) and T4 (penguin-like) topologies.
For T1 and T4 topologies, there exist different  possibilities of assigning the fields in $\mathcal{O}_5$ to the four external legs, hence several one-loop diagrams appear, see Fig. \ref{fig:T1-topology}. 
The mediator fields in all the one-loop diagrams are denoted  by $X_i$, where $i$ runs from 1 to 3(4) for the  T3 topology (T1 and T4 topologies) and denotes the number of mediators. The Lorentz nature, scalar or fermion, is shown in the corresponding one-loop diagram by a dashed or solid line, respectively. 

In Table \ref{tab:charges} we show the L and PQ charges of the mediator fields that are consistent with our assumptions: $i)$ axionic one-loop realizations of the $\mathcal{O}_5$ operator and prohibition of the $\mathcal{O}_4$ operator, $ii)$ lepton number conservation, $iii)$ vanishing contributions of the tree-level realizations of $\mathcal{O}_5$ to the neutrino mass matrix, and $iv)$ existence of a remnant $Z_N$ symmetry after the PQ breaking (see next Section). 
For completeness purposes we display in the Appendix \ref{sec:app-models} the sets of quantum numbers of the mediator fields for each model~\cite{Yao:2018ekp}.
It turns out that for $\alpha=0, 1$ or 2 (the values leading to either  $Y=0$ singlet fermions, or $Y=\pm1$ doublet fermions or $Y=\pm1$ doublet scalars) one or two Dirac seesaw mechanisms arise, implying that the one-loop contribution is subdominant (the exceptional case is the T4-3-I-V model since all the mediators fields are $SU(2)_L$ triplets).

\begin{itemize}
\item {\bf T1 models.} The six T1 models have four mediators, where the T1-1-A and T1-1-B models have  three scalar mediators, T1-3-D and T1-3-E models have two scalars, whereas there is one single scalar mediator in T1-2-A and T1-2-B models. The  T1 models differ on the number of scalar and fermion mediator fields, or on the exchange of the $S$ and $H$.  Note that models  T1-1-A and T1-2-A share the same L and PQ charges, and this is the same for T1-1-B and T1-2-B models. This is because the charge assignment is independent of the $SU(2)_{L}$ transformation properties of the fields. 
Each T1 model has four sets of possible electroweak charges for the mediator fields, all of them featuring at least one of the mediators of the Dirac seesaw mechanisms when  $\alpha=0, 1$ or 2 (see Table \ref{tab:T1-quantum_numbers}). 
\item {\bf T3 models.} There is one type of model with a T3 topology, which involves three mediators with two of them scalars. 
The T3 model has four sets of possible electroweak charges for the mediator fields, with at least one of the mediators of the Dirac seesaw mechanisms when  $\alpha=0, 1$ or 2 (see Table \ref{tab:T3-quantum_numbers}). 
\item {\bf T4 models.}
The T4-3-I models have four mediators with two fermions, with the particular feature that $X_1$ is not running inside the loop. This in turn implies that all its quantum numbers are fixed: it transforms as a $SU(3)_L$ triplet with zero hypercharge and the L and PQ charges are settled by the lepton doublet charges:  L$(X_1)=-1$ and $\mathcal{X}_{X_1}=-\mathcal{X}_L=-(\mathcal{X}_{N_R}+\mathcal{X}_S)$. 
On the other hand, the T4-3-I-V model does not contain any mediator field of the Dirac seesaw mechanisms, however, it requires the mixing between the neutral and charged leptons with the neutral and charged components of $X_1$, which induce charged lepton flavor violating processes at tree level. 
\end{itemize}

\begin{table}[t!]
\begin{centering}
\begin{tabular}{|c||c||c|c|c|c|}
\hline 
 Model & Symmetry & $X_1$  & $X_2$  & $X_3$  & $X_4$   \tabularnewline
\hline 
\hline 
\multirow{2}{1.7cm}{\centering T1-(1/2)-A} 
      &$U(1)_\textrm{L}$  & $\delta-1 $  & $\delta $  & $\delta $  & $\delta $  \\\cline{2-6}
      &$U(1)_\textrm{PQ}$  & $\mathcal{X}_{X_4}-(\mathcal{X}_{N_R}+\mathcal{X}_S) $  & $\mathcal{X}_{X_4}-\mathcal{X}_S $  & $\mathcal{X}_{X_4} $  & $\mathcal{X}_{X_4} $  \\\cline{2-6}
  \hline
  \hline 
\multirow{2}{1.7cm}{\centering T1-(1/2)-B} 
      &$U(1)_\textrm{L}$  & $\delta-1 $  & $\delta $  & $\delta $  & $\delta $  \\\cline{2-6}
      &$U(1)_\textrm{PQ}$  & $\mathcal{X}_{X_4}-(\mathcal{X}_{N_R}+\mathcal{X}_S) $  & $\mathcal{X}_{X_4}-\mathcal{X}_S $  & $\mathcal{X}_{X_4}-\mathcal{X}_S $  & $\mathcal{X}_{X_4} $  \\\cline{2-6}
  \hline
  \hline 
\multirow{2}{1.7cm}{\centering T1-3-D} 
      &$U(1)_\textrm{L}$  & $\delta $  & $\delta+1 $  & $\delta+1 $  & $\delta $  \\\cline{2-6}
      &$U(1)_\textrm{PQ}$  & $\mathcal{X}_{X_4} $  & $\mathcal{X}_{N_R}+\mathcal{X}_{X_4} $  & $\mathcal{X}_{{N_R}}+\mathcal{X}_{X_4}+\mathcal{X}_S $  & $\mathcal{X}_{X_4} $  \\\cline{2-6}
  \hline
  \hline 
\multirow{2}{1.7cm}{\centering T1-3-E} 
      &$U(1)_\textrm{L}$  & $\delta $  & $\delta+1 $  & $\delta+1 $  & $\delta $  \\\cline{2-6}
      &$U(1)_\textrm{PQ}$  & $\mathcal{X}_{X_4}+\mathcal{X}_S $  & $\mathcal{X}_{{N_R}}+\mathcal{X}_{X_4}+\mathcal{X}_S $  & $\mathcal{X}_{N_R}+\mathcal{X}_{X_4}+\mathcal{X}_S $  & $\mathcal{X}_{X_4} $  \\\cline{2-6}
\hline 
\hline 
\multirow{2}{1.7cm}{\centering T3-1-A} 
      &$U(1)_\textrm{L}$  & $\delta $  & $\delta+1 $  & $\delta+1 $ &   \\\cline{2-5}
      &$U(1)_\textrm{PQ}$  & $\mathcal{X}_{X_1} $  & $\mathcal{X}_{{N_R}}+\mathcal{X}_{X_1}$  & $\mathcal{X}_{{N_R}}+\mathcal{X}_{X_1}+\mathcal{X}_S $  & \\\cline{2-5}
\hline
\hline
\multirow{2}{1.7cm}{\centering T4-3-I} 
      &$U(1)_\textrm{L}$  & $-1 $  & $\delta-1 $  & $\delta $  & $\delta $  \\\cline{2-6}
      &$U(1)_\textrm{PQ}$  & $-(\mathcal{X}_{N_R}+\mathcal{X}_S) $  & $\mathcal{X}_{X_4}-(\mathcal{X}_{N_R}+\mathcal{X}_S) $  & $\mathcal{X}_{X_4}-\mathcal{X}_S $  & $\mathcal{X}_{X_4} $  \\\cline{2-6}
\hline
\end{tabular}
\par\end{centering}
\caption{L and PQ charges of the mediator fields $X_i$ of T1, T3 and T4 models. 
$\delta$, $\mathcal{X}_{X_1}$ and $\mathcal{X}_{X_4}$ are free real parameters whereas $\mathcal{X}_S$ and $\mathcal{X}_{N_R}$ denote the PQ charges of $S$ and $N_R$, respectively. The Lorentz nature of $X_i$ is inferred from the corresponding one-loop diagram (see text for details).}
\label{tab:charges}
\end{table}

In the next Section we will present the conditions on the charges of the mediator fields that must be fulfilled to guarantee, at the same time, the presence of a WIMP candidate in the  spectrum and the absence of tree-level Dirac seesaw mechanisms.

\section{Radiative neutrino masses and Axion/WIMP dark matter}
It turns out that the same values for $\alpha$ that lead to tree-level Dirac seesaw mechanisms are the ones that allow us to have an electrically neutral and PQ charged particle in the spectrum, thus opening the possibility of having multicomponent DM scenarios comprising such particle (a WIMP) and the axion. 
Concretely, the existence of a neutral particle in the spectrum is guaranteed by setting $Q=T_3+Y/2=0$ such that $Y=-2T_3$ for at least one of the mediators.  This implies that $\alpha$ can only take the values 0, 1 or 2 (depending upon the $SU(2)_L$ representation), which are precisely the same values  that lead to the tree-level realizations of $\mathcal{O}_5$. Thus, the presence of a WIMP candidate in the models automatically implies that the one-loop contribution is the main contribution to the neutrino masses. 

Regarding the stability of the WIMP particle, all the mediator fields (except $X_1$ in T4 models\footnote{It is worth mentioning that in the T4 models the $X_1$ field is not running inside the loop which makes it different from the other mediator fields in what concerns  DM.}) can be made part of a dark (WIMP) sector by demanding that a residual $Z_N$ symmetry is generated after the $U(1)_\textrm{PQ}$ breaking, in the same line of the scotogenic models.
This is achieved  when $\mathcal{X}_S$ is an even integer $N$ (to preserve the periodicity of the axion potential) and  the PQ charges of the loop mediators are  $\mathcal{X}_{X_i}\neq0 $ mod$(\mathcal{X}_S)$. 

On the other hand,  the conditions over the PQ charges of the mediator fields that avoid tree-level Dirac seesaw mechanisms (and hence guarantee a WIMP particle in the spectrum) lead to PQ charges $\mathcal{X}_{X_i}$ of the form of (see Table \ref{tab:charges})
\begin{align}\label{conditions PQ charges of the mediator}
  &\mathcal{X}_{X_4},\,\, \mathcal{X}_{X_4}\pm \mathcal{X}_S,\,\, \mathcal{X}_{X_4}\pm \mathcal{X}_{N_R},\,\, \textrm{or}\,\, \mathcal{X}_{X_4}\pm \mathcal{X}_{N_R}\pm \mathcal{X}_S; \hspace{1cm}(\textrm{for T1 and T4 models}),\nonumber\\
&    \mathcal{X}_{X_1},\,\, \mathcal{X}_{X_1}\pm \mathcal{X}_{N_R},\,\, \textrm{or}\,\, \mathcal{X}_{X_1}\pm \mathcal{X}_{N_R}\pm \mathcal{X}_S; \hspace{2.8cm}(\textrm{for T3 models}). 
\end{align}
Consequently, in order to ensure a non-trivial charge for these fields under the remnant $Z_N$ ($N=\mathcal{X}_S$), the $\mathcal{X}_{X_4}$ charge ($\mathcal{X}_{X_1}$ for T3-1-A model) must be an integer non-multiple of $\mathcal{X}_S$ as long as $N_R$ transforms trivially, $\mathcal{X}_{N_R}=0$ mod$(\mathcal{X}_S)$ (this the simplest choice, however, see Sec.~\ref{sec. particular case} for other possible $Z_N$ assignment).
Note further that if $\mathcal{X}_{N_R}\neq0$ and $2\mathcal{X}_{N_R}\neq\mathcal{X}_{S}$ the L conservation would not be mandatory since the Diracness of neutrinos would be guarantee for the $Z_N$ properties of the lepton doublet, $L\,\cancel{\sim}\,\omega^{N/2}$ (see Ref. \cite{Hirsch:2017col} for details and and Sec. \ref{sec. particular case} for an example). 

To illustrate a specific charge assignment, let us assume $\mathcal{X}_S=2$ and $\mathcal{X}_{N_R}=0$, which entails that the PQ charge of the lepton doublet and singlet are both 2 (this is because of the Yukawa terms $\bar{L}\tilde{H}N_RS$ and $\bar{L}H\ell_R$). Thus, if all the mediators running inside the loop have odd PQ charges then a remnant $Z_2$ symmetry would appear in such a way that they are $Z_2$ odd while the rest of the  model particles are $Z_2$ even.

The stability of electrically neutral particles naturally leads to WIMP scenarios.  Among the possible scenarios that may arise (see Tables \ref{tab:T1-quantum_numbers}, \ref{tab:T3-quantum_numbers} and \ref{tab:T4-quantum_numbers}) we highlight the ultraviolet fermionic realizations of the Higgs portal \cite{Patt:2006fw}: the singlet (via the axion interactions),  singlet-doublet  and doublet-triplet Dirac DM \cite{Kim:2008pp,Freitas:2015hsa,Yaguna:2015mva,Bhattacharya:2015qpa}; and the renormalizable scalar DM models: the singlet scalar DM model \cite{Silveira:1985rk,McDonald:1993ex,Burgess:2000yq}, inert doublet model (IDM) \cite{Deshpande:1977rw,Barbieri:2006dq} and inert triplet model \cite{Cirelli:2005uq,FileviezPerez:2008bj,Hambye:2009pw}, along with the interplays singlet-doublet and doublet triplet \cite{Kadastik:2009dj,Kadastik:2009cu,Kakizaki:2016dza,Liu:2017gfg,Betancur:2017dhy}. 

Regarding the heavy quark $D$,  since the SM quarks and $H$ are not charged under the PQ symmetry, $D_R$ cannot couple to the SM quark doublet through $\overline{q_L}HD_R$ as long as $\mathcal{X}_{D_R}\neq0$.
However, it may interact with the quark singlet via the $\overline{D_L}d_RS$ term, which entails that $\mathcal{X}_{D_L}\neq \mathcal{X}_{S}$ to avoid the constraints due to the mixing the SM quarks.  With respect to the interactions with the dark sector, if there exists a $Y=0 $ singlet scalar $\varphi$ or a second Higgs doublet $H_2$ in the spectrum then the terms $\overline{D_L}d_R\varphi$ and $\overline{q_L}H_2D_R$ would appear.
The first term would require that $\mathcal{X}_{D_L}=\mathcal{X}_{\varphi}$ while the second one $\mathcal{X}_{D_R}=-\mathcal{X}_{H_2}$, so $D$ must be also part of the dark sector in order for such terms be allowed.
It follows that $D$ can only decay into dark scalars ($\varphi$ and/or $H_2$) and SM quarks, leading to SUSY-like signals such as jets plus missing energy at the LHC if $D$ is not heavy enough (see {\it e.g.} Ref.~\cite{Alves:2016bib} for an explicit analysis). 
In the opposite case, the Lagrangian would be invariant under the exchange $(D_L,D_R)\rightarrow(-D_L,-D_R)$ making it a stable colored heavy quark, thus bringing some cosmological problems \cite{Nardi:1990ku}. Therefore, the instability of the heavy quark $D$ imposes additional constraints over the viable models since it demands the existence of either a $Y=0 $ singlet scalar or an  extra Higgs doublet in the dark sector. From Tables \ref{tab:T1-quantum_numbers}, \ref{tab:T3-quantum_numbers} and \ref{tab:T4-quantum_numbers}, we obtain that the following models do not contain such fields: T1-2-A-IV, T1-3-D-IV and T4-3-I-V.  

On the other hand, since the PQ mechanism is at work suitable parameters for the QCD axion can be considered in order to have the axion as a second DM candidate. In the framework of a broad class of inflationary scenarios QCD axions may be non-thermally produced in the early Universe through the vacuum-realignment mechanism and surviving as cold DM matter. The axion contribution to energy density  depends on the order in which cosmological events take place, especially whether the breaking of the PQ symmetry occurred before or after inflation.  For the case of reheating temperatures lower than $f_a$, the axion relic density due to the misalignment population is fixed by the initial field displacement, the so-called misalignment angle $\theta_a$, and $f_a$ \cite{Abbott:1982af,Bae:2008ue},
\begin{equation}
 \Omega_ah^2\approx 0.18\theta_a^2\left(\frac{f_a}{10^{12} \textrm{ GeV}}\right)^{1.19}.
\end{equation}
Thus the allowed window of axion parameters is
\begin{align}
  10^9\,\textrm{GeV}&\lesssim f_a\lesssim \theta_a^{-2}\,10^{12} \,\textrm{GeV},\\
    6\,\theta_a^{2} \,\mu\textrm{eV}&\lesssim m_a\lesssim 6\,\textrm{meV}
\end{align}
where $m_a\approx6\mu\,\textrm{eV}(10^{12}\,\textrm{GeV}/f_a)$ \cite{Wilczek:1977pj,Weinberg:1977ma} (see Refs.~\cite{DiLuzio:2017pfr,DiLuzio:2016sbl} for the detailed study of the window for preferred axion models and Refs. \cite{Graham:2015ouw,Du:2018uak,Brubaker:2016ktl,Akerib:2017uem,Fu:2017lfc} for the current status of experimental searches). For this case the axion population from the decay of topological defects such as string axions and domain walls is diluted by inflation. 

It follows that in this framework, the DM of the Universe comprises the WIMP and the axion, which behave as two completely independent DM particles, without affecting the standard relic density calculations and the corresponding experimental bounds ~\cite{Baer:2011hx,Bae:2013hma,Dasgupta:2013cwa,Alves:2016bib,Ma:2017zyb,Chatterjee:2018mac,Betancur:2018xtj}.
Therefore, at the electroweak scale the models presented in this work are up to some extent similar to the renormalizable WIMP DM models mentioned above, with the main difference being that the DM is complemented by axions.  And since there exist two contributions to the total DM relic abundance, we expect that all the constraints over such WIMP scenarios would be substantially weakened, thus resulting  in larger portions of the parameter space that are still allowed within those scenarios (see Refs.~\cite{Dasgupta:2013cwa,Alves:2016bib,Betancur:2018xtj,Chatterjee:2018mac} for specific scenarios). 
In addition to this, since the WIMP may couple to the right-handed neutrino through Yukawa interactions further WIMP annihilation channels may appear and thus changing drastically the expected DM phenomenology,  with the bonus that they are rather unconstrained because such interactions do not induce lepton flavor violation processes.  

\section{A case study}\label{sec. particular case}
In this section we study a specific  model resulting from the analysis made in previous sections, the T3-1-A-I model with $\alpha=0$ and a residual $Z_2$ symmetry.
Thus, the model contains the following fields:  an $SU(2)_L$ doublet scalar $X_3$, and two SM singlets, one being a Dirac fermion $X_1$ while the other one is a complex scalar $X_2$.
We add a second singlet fermion in order to obtain two massive neutrinos.
In what follows we rename these fields as $X_1\to\psi^c$, $X_2\to \varphi^*$ and $X_3\to \tilde{H}_2$, with $\varphi=(\varphi_{R}+i\varphi_{I})/\sqrt{2}$, $H_2=(H^+,(H^0+iA^0)/\sqrt{2}))^T$ and $\tilde{H}_2=i\sigma_2H_2^*$. 
Moreover, $H_1$ will denote the SM Higgs doublet. 
The $U(1)_\textrm{PQ}$ and $U(1)_\textrm{L}$ charges are given in Table \ref{tab:chargesT31}. 

\begin{table}[t!]
  \setlength{\tabcolsep}{5pt} \global\long\def\arraystretch{1.1}
  \centering
  \begin{tabular}{|c|c|c|c|c|c|c|c|c|c|}
    \hline
    & $L$ & $\ell_R$ & $N_R$ & $S$  &  $\psi$  & $\varphi$    & $H_2$       & $D_L$   & $D_R$\\
    \hline
    \hline
    $U(1)_\textrm{L}$  & $1$ & $1$   & $1$   & $0$  &  $1$         & $0$      & $0$         & $0$    & $0$ \\
    \hline
    $U(1)_\textrm{PQ}$  & $2$ & $2$   & $0$   & $2$  &  $3$         & $3$      & $1$         & $1$    & $-1$ \\
    \hline
    $Z_2$     & $+$ & $+$   & $+$   & $+$  & $-$         & $-$      & $-$         & $-$    & $-$ \\
    \hline   
  \end{tabular}
  \caption{L and PQ charges for model T3-1-A-I with $\alpha=0$ and $\delta=-1$, $\mathcal{X}_S=2$, $\mathcal{X}_{\psi}=1$ and $\mathcal{X}_{N_R}=0$. The charges under the remnant $Z_2$ symmetry are also shown.}
  \label{tab:chargesT31}
\end{table}

The more general Lagrangian invariant under $\mathcal{G}_\textrm{SM}\times U(1)_\textrm{L}\times U(1)_\textrm{PQ}$ symmetry comprises the interaction terms $   -\mathcal{L}\supset\mathcal{V}_{H_1,S}+\mathcal{V}_\varphi+\mathcal{V}_{H_2}+\mathcal{V}_1+\mathcal{L}_1$, 
 where
 \begin{align}
   \mathcal{V}_{H_1,S}&=-\mu_1^2|H_1|^2+\lambda_1|H_1|^4+\mu^2_S|S|^2+\lambda_S|S|^4+\lambda_{6} |H_1|^2|S|^2,\\
   \mathcal{V}_\varphi&=\mu^2_\varphi|\varphi|^2+\lambda_\varphi|\varphi|^4+\lambda_7 |H_1|^2|\varphi|^2+\lambda_{8} |\varphi|^2|S|^2,\\
   \mathcal{V}_{H_2}&=\mu^2_2|H_2|^2+\lambda_2|H_2|^4+\lambda_3 |H_1|^2|H_2|^2+\lambda_4 |H_1^\dagger H_2|^2+\lambda_9 |H_2|^2|S|^2,\\
   \mathcal{V}_1&=\lambda_{10} |H_2|^2|\varphi|^2+\lambda_{11}\left[ \varphi^* S  \tilde{H}_2^\dagger  \tilde{H}_1 +\textrm{h.c.}\right],\\
   \mathcal{L}_1&= y_{i\beta}\overline{\psi_i}\tilde{H}_2^\dagger L_{\beta}+h_{\beta i}\varphi^*\overline{N_{R\beta}} \psi_i+y_Q S 
   \bar{D}_LD_R+f_\beta\overline{q_{L\beta}}H_2D_R+\textrm{h.c.}.\label{Yukawa Lagrangian}
 \end{align}
Here some comments are in order. The $\lambda_6$ term mixes the real component of the axion with the Higgs field, so we can neglect the $\lambda_6$ without lost of generality. 
$\lambda_{11}$ is responsible for the mixing between the neutral components of $\varphi$ and $H_2$ that leads to Dirac neutrino masses (notice that $\lambda_{11}$ plays, to some extent, the role of the $\lambda_5$ term in the scotogenic model \cite{Ma:2006km}). 
Since the SM quarks are not charged under the PQ symmetry, $D$ does not couple to SM particles through $\overline{D_L}d_R\varphi$ but does through $\overline{q_L}H_2D_R$. 
Note that, under the $Z_2$ remnant symmetry\footnote{In fact, a larger dark $U(1)$ symmetry is obtained, which contains the remnant $Z_2$ as a subgroup.} the fields running in the loop of diagram T3-1-A in Fig. \ref{fig:T1-topology} are odd. Therefore, the lightest of them can be considered as the DM candidate (if it is neutral). This model at low energies, {\it i.e.}, without the axion, is identical to the one presented in Ref.~\cite{Farzan:2012sa}, since instead of considering an ultraviolet completion realization it introduced a soft breaking term that allows the mixing between the neutral components of the singlet and doublet (in a similar way as $\lambda_{11}$ does). 

Let us consider the limit $\lambda_6\rightarrow 0$ in order to avoid the mixing between the scalar $S$ and the SM Higgs boson, and the limit $\lambda_{8,9}\rightarrow 0$ with the aim to avoid a large fine-tuning in the masses of $\varphi$ and $H^0$.  In the basis $\big(H^0, \varphi_R \big)^T$ and $(A^0, \varphi_I)^T$ the mass matrices for the $Z_2$-odd neutral scalars are given by
\begin{equation}\label{Meven}
 \mathcal{M}_{R,I}=
 \left( 
  \begin{array}{cc}
    \mu_2^2+\frac{1}{2}(\lambda_3+\lambda_4)v^2 & \frac{1}{2}\lambda_{11} v_Sv\\
   \frac{1}{2} \lambda_{11}v_Sv& \mu_\varphi^2+\frac{1}{2}\lambda_7v^2
  \end{array}
 \right ), \hspace{1cm}\sin(2\theta_{R,I})=\frac{\lambda_{11} v v_S}{m^2_{S_{R,I2}}-m^2_{S_{R,I1}}},
\end{equation}
where 
$m_{S_{Rj}}$ ($m_{S_{Ij}}$) are the two mass eigenvalues of $\mathcal{M}_{R}$ ($\mathcal{M}_{I}$).  The mass of the charged scalar $H^+$ is given by $m^2_{H^+}=\mu_2^2+\frac{1}{2}\lambda_3v^2$, as in the IDM. 

Neutrino masses are generated at one-loop through the Feynman diagram corresponding to the T3-1-A model in Fig. \ref{fig:T1-topology}.  The effective mass matrix is given by
\begin{equation}\label{neutrino masses}
 \big(M_\nu\big)_{\beta\beta'}=\frac{1}{64\pi^2}\frac{ \lambda_{11}v_Sv}{m_{S_{R2}}^2-m_{S_{R1}}^2}\sum_ih_{\beta i}y_{ i\beta'}m_{\psi_i}\left[F\left(\frac{m_{S_{R2}}^2}{m_{\psi_i}^2}\right)-F\left(\frac{m_{S_{R1}}^2}{m_{\psi_i}^2}\right)\right]+(R\rightarrow I),
\end{equation}
where $F(x)=x\ln(x)/(x-1)$, and $R\rightarrow I$ means the exchange of the CP-even eigenstates ($S_{Rj}$) with the corresponding CP-odd eigenstates ($S_{Ij}$).
It follows that the neutrino masses are suppressed by several factors: the loop factor, the Yukawa couplings, the mass of the heaviest loop mediator and  $\lambda_{11}$. It turns out that the resulting suppression is enough to counteract the contribution coming from $v_S$ (see Eq. (\ref{eq:numerical-estimate})). 
A simple numerical estimate for the effective mass matrix can be calculated considering the limit\footnote{The limits $\lambda_{3,4,7}\to 0$ are motivated by the naturalness of the scalar potential since they avoid the imposition of large cancellations among the scalar potential parameters.} $\lambda_{3,4,7}\rightarrow0$, $\mu_2^2\rightarrow\mu_\varphi^2$ and $\mu_\varphi^2\gg \lambda_{11}v_Sv$. In this case,
\begin{align}
 \big(M_\nu\big)_{\beta\beta^\prime}&\approx\frac{\lambda_{11} v_S v}{32\pi^2}\sum_ih_{\beta i}y_{i\beta^\prime}\frac{m_{\psi_i}}{\mu_\varphi^2-m_{\psi_i}^2}\Bigg[1-\frac{m_{\psi_i}^2}{\mu_\varphi^2-m_{\psi_i}^2}\log\Bigg(\frac{\mu_\varphi^2}{m_{\psi_i}^2}\Bigg)\Bigg].
\end{align}
If $\mu_\varphi^2\gg m_\psi^2$, that is considering fermion DM, it reduces to 
\begin{align}\label{eq:numerical-estimate}
  \big(M_\nu\big)_{\beta\beta^\prime}&\approx\frac{\lambda_{11} v_S v}{32\pi^2}\sum_ih_{\beta i}y_{i\beta^\prime}\frac{m_{\psi_i}}{\mu_\varphi^2}\nonumber\\
 &\sim0.05\textrm{ eV}\times\Bigg(\frac{h_{\beta i}y_{i\beta^\prime} }{10^{-4}}\Bigg)\Bigg(\frac{\lambda_{11}}{10^{-10}}\Bigg)\Bigg(\frac{m_{\psi_i}}{10^2 \textrm{ GeV}}\Bigg)\Bigg(\frac{5\times10^3\textrm{ GeV}}{\mu_\varphi}\Bigg)^2\Bigg(\frac{v_S}{10^{9}\textrm{ GeV}}\Bigg).
\end{align}
The linear dependence of the neutrino masses with $v_S$ arises because the loop mediators are not lying at the PQ scale, a feature that is also shared with other radiative neutrino mass models e.g.,  \cite{Bertolini:2014aia}. 
Note that the smallness of $\lambda_{11}$ is natural in the ’t Hooft sense \cite{tHooft:1979rat} since $\lambda_{11}=0$ leads to an extra symmetry (the charge of $S$ is no longer connected with the charges of the other particles). 
Furthermore, the smallness of $\lambda_{11}$ is a necessary condition in order to have the loop scalar mediators at or below the TeV mass scale. 
This is also in consonance with the requirement of demanding a tiny value  for the scalar coupling between the Higgs boson $H_1$ and the axion field $S$ (these conditions are unavoidable in axion models as long as light loop mediators are required \cite{Dasgupta:2013cwa}). 

\begin{figure}
  \centering
 \includegraphics[scale=0.5]{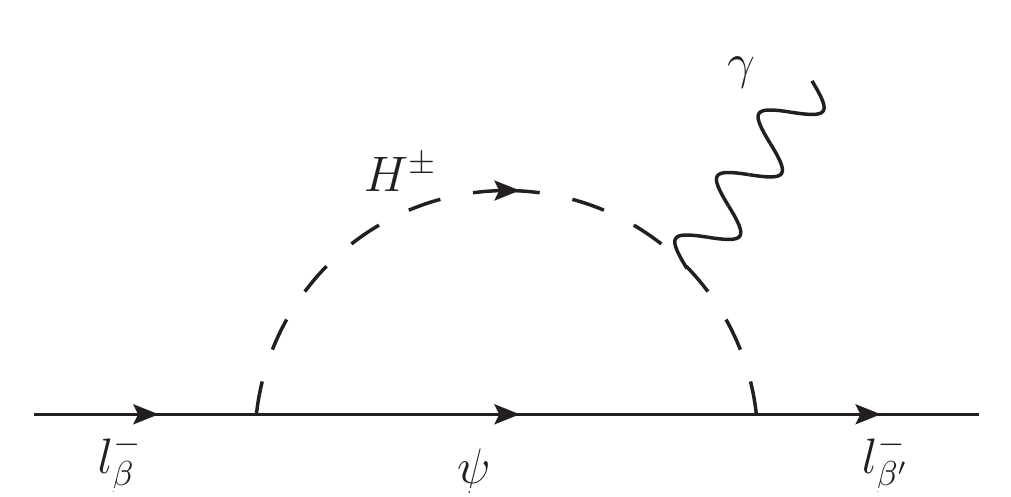}
    \caption{Feynman diagram for the process $\ell_\beta\rightarrow \ell_{\beta^\prime}\gamma$.}
  \label{LFV charged}
 \end{figure}

On the other hand, the Yukawa interactions in Eq.~(\ref{Yukawa Lagrangian}) also lead to lepton flavor violation (LFV) processes at one-loop mediated by the charged scalar $H^\pm$ and $\psi_i$ (see Fig.~\ref{LFV charged}).
The decay rate for this process is given by 
\begin{equation}
 \Gamma\big(\ell_\beta\rightarrow \ell_{\beta^\prime}\gamma\big)=\frac{e^2m_\beta^5}{16\pi}\frac{1}{16^2\pi^4m^4_{H^\pm}}\sum_i\left|y_{i\beta}y_{i\beta^\prime}^*\right|^2\left(\frac{2t_i^2+5t_i-1}{12(t_i-1)^3}-\frac{t_i^2\log t_i}{2(t_i-1)^4}\right)^2.
\end{equation}
where $t_i\equiv m_{\psi_i}^2/m_{H^\pm}^2$. Considering the limit of the heavy scalar, $m_{H^\pm}\gg m_{\psi_i}$, it follows that
\begin{equation}
 \left|y_{i\mu}y_{i e}^*\left(\frac{5\times10^3\textrm{ GeV}}{m_{H^\pm}}\right)^2\right|\lesssim0.1,
\end{equation}
where the bound $\textrm{BR}(\mu\rightarrow e\gamma)<5.7\times10^{-13}$ \cite{Adam:2013mnn} has been used. 

Regarding WIMP DM, this model features either singlet fermion or singlet/doublet scalar candidates\footnote{A detailed DM phenomenological study of this model will be done elsewhere.}.  
Such a WIMP-like setup is similar to the model studied in Ref.~\cite{Farzan:2012sa} with the difference that in the present case the DM is additionally composed by axions. 
Therefore, all the constraints that may exist on the model with only WIMPs would be substantially weakened. 
Indeed, when the WIMP particle is mainly doublet (an IDM-like scenario) most of the intermediate DM mass range can be free of DM constraints \cite{Dasgupta:2013cwa,Alves:2016bib} (see Ref.~\cite{Chatterjee:2018mac} for the case of singlet scalar DM).
In the case of fermion DM, in order to account for the total relic abundance the $\bar{\psi}\psi$ annihilation must be driven by $h_{\beta i}$ Yukawa terms, since in this case there are not constraints coming from lepton flavor violating processes, so they can be large enough \cite{Farzan:2012sa}. 
In contrast,  $\bar{\psi}\psi$ annihilation via  $y_{i\beta }$ Yukawa terms is not a viable option without some detailed fine-tuning of parameters. All in all, this shows the impact of considering mixed axion/WIMP DM scenarios. 

Finally, it is worth mentioning that with a suitable PQ charge assignment (such as the one displayed in Table \ref{tab:chargesT31-2}) and without imposing lepton number conservation it is possible to obtain a consistent model where the Diracness of neutrinos is guaranteed. 
In this case, a remnant $Z_4$ symmetry is obtained from the PQ symmetry breaking which is responsible for the stability the WIMP candidate. 

\begin{table}[h!]
  \setlength{\tabcolsep}{5pt} \global\long\def\arraystretch{1.1}
  \centering
  \begin{tabular}{|c|c|c|c|c|c|c|c|c|c|}
    \hline
    & $L$ & $\ell_R$ & $N_R$ & $S$  &  $\psi$  & $\varphi$    & $H_2$       & $D_L$   & $D_R$\\
    \hline
    \hline
    $U(1)_\textrm{PQ}$  & $8$ & $8$   & $4$   & $4$  &  $7$         & $3$      & $-1$         & $5$    & $1$ \\
    \hline
    $Z_4$     & $1$ & $1$   & $1$   & $1$  & $\omega^3$         & $\omega^3$      & $\omega^3$         & $\omega$    & $\omega$ \\
    \hline   
  \end{tabular}
  \caption{Alternative PQ charges for model T3-1-A-I that guarantee the Diracness of neutrinos without further assuming the lepton number symmetry. The charges under the remnant $Z_4$ symmetry are also shown, where $\omega^4=1$.}
  \label{tab:chargesT31-2}
\end{table}

\section{Summary}
In this work we have studied the one-loop realizations of the $d=5$ operator  $\bar{L}\tilde{H}N_RS$ that leads to Dirac neutrino masses, with $S$ being a singlet scalar field that hosts the QCD axion.
As usual, the axion arises from the breaking of the Peccei-Quinn symmetry, which in our setup we used to not  only solve the strong CP problem, but also to forbid the operator $\overline{L}\widetilde{H}N_{R}$ (which generates Dirac neutrino masses at tree level) and the tree-level realizations of $\bar{L}\tilde{H}N_RS$. Thus, the neutrino masses are directly correlated to the axion mass  (via the PQ symmetry breaking scale $v_s$) and their smallness is due to the radiative character besides the mass suppression of the loop mediators.   Furthermore, the PQ symmetry breaking leaves  a residual $Z_N$ symmetry that allow us to guarantee the stability of the lightest of the mediators in the one-loop neutrino mass diagrams (as happens in the scotogenic models), thus leading naturally to  multicomponent DM scenarios with axions and WIMPs.  We have illustrated our proposal by considering a specific model, where simple numerical estimates allow us to show the effectiveness of the scheme regarding neutrino masses, DM and lepton flavor violating processes.

\section{ACKNOWLEDGMENTS}
We are thankful to Diego Restrepo for enlightening discussions.  
Work supported by Sostenibilidad-UdeA and the UdeA/CODI Grant 2017-16286, and by COLCIENCIAS through the Grants 111565842691 and 111577657253. C. D. R. C. acknowledges the financial support given by the Departamento Administrativo de Ciencia, Tecnología e Innovación - COLCIENCIAS (doctoral scholarship 727-2015). O.Z. acknowledges the ICTP Simons associates program and the kind hospitality of the Abdus Salam ICTP where part of this work was done.

\section*{Appendix}\label{sec:app-models}
Below  we display the sets of quantum numbers of the mediator fields for each model~\cite{Yao:2018ekp}.
 \begin{table}[h]
 \begin{subtable}[b]{0.5\textwidth}
 \setlength{\tabcolsep}{0.15cm} \global\long\def\arraystretch{1.2}
 \begin{centering}
 \begin{tabular}{|c||c||c|c|c|c||c|}
 \hline 
  Model & Sol. & $X_1^F$  & $X_2^S$  & $X_3^S$  & $X_4^S$ & $\alpha$  \tabularnewline
 \hline 
 \hline 
 \multirow{4}{1.7cm}{\centering T1-1-A} 
        & I   & $1_\alpha $  & $1_\alpha $  & $1_\alpha  $  & $2_{\alpha-1}$ & $0,\ 2$\\\cline{2-7}
        & II   & $2_\alpha $  & $2_\alpha $  & $2_\alpha  $  & $1_{\alpha-1}$ & $\pm1$\\\cline{2-7}
        & III   & $2_\alpha $  & $2_\alpha $  & $2_\alpha  $  & $3_{\alpha-1}$ & $\pm1$\\\cline{2-7}
        & IV   & $3_\alpha $  & $3_\alpha $  & $3_\alpha  $  & $2_{\alpha-1}$ & $0,\ 2$\\\cline{2-7}
 \hline 
 \multirow{4}{1.7cm}{\centering T1-1-B} 
        & I   & $1_\alpha $  & $1_\alpha $  & $2_{\alpha-1}  $  & $2_{\alpha-1}$ & $0,\ 2$\\\cline{2-7}
        & II   & $2_\alpha $  & $2_\alpha $  & $1_{\alpha-1}  $  & $1_{\alpha-1}$ & $\pm1$\\\cline{2-7}
        & III   & $2_\alpha $  & $2_\alpha $  & $3_{\alpha-1}  $  & $3_{\alpha-1}$ & $\pm1$\\\cline{2-7}
        & IV   & $3_\alpha $  & $3_\alpha $  & $2_{\alpha-1}  $  & $2_{\alpha-1}$ & $0,\ 2$\\\cline{2-7}
        \hline
 \end{tabular}
 \par\end{centering}
 \protect\caption{ }
 \label{tab:T1(12)A_QN}
 \end{subtable}
 \begin{subtable}[b]{0.5\textwidth}
 \setlength{\tabcolsep}{0.15cm} \global\long\def\arraystretch{1.2}
 \begin{centering}
 \begin{tabular}{|c||c||c|c|c|c||c|}
 \hline 
  Model & Sol. & $X_1^S$  & $X_2^F$  & $X_3^F$  & $X_4^F$ & $\alpha$  \tabularnewline
 \hline 
 \hline 
 \multirow{4}{1.7cm}{\centering T1-2-A} 
       & I  & $1_\alpha $  & $1_\alpha $  & $1_\alpha  $  & $2_{\alpha-1} $ & $0,\ 2$ \\\cline{2-7}
       & II  & $2_\alpha $  & $2_\alpha $  & $2_\alpha  $  & $1_{\alpha-1} $ & $\pm1$ \\\cline{2-7}
       & II  & $2_\alpha $  & $2_\alpha $  & $2_\alpha  $  & $3_{\alpha-1} $ & $\pm1$ \\\cline{2-7}
       & IV  & $3_\alpha $  & $3_\alpha $  & $3_\alpha  $  & $2_{\alpha-1} $ & $0,\ 2$ \\\cline{2-7}
 \hline
 \multirow{4}{1.7cm}{\centering T1-2-B} 
       & I  & $1_\alpha $  & $1_\alpha $  & $2_{\alpha-1} $  & $2_{\alpha-1}$ & $0,\ 2$\\\cline{2-7}
       & II  & $2_\alpha $  & $2_\alpha $  & $1_{\alpha-1} $  & $1_{\alpha-1}$ & $\pm1$\\\cline{2-7}
       & III  & $2_\alpha $  & $2_\alpha $  & $3_{\alpha-1} $  & $3_{\alpha-1}$ & $\pm1$\\\cline{2-7}
       & IV  & $3_\alpha $  & $3_\alpha $  & $2_{\alpha-1} $  & $2_{\alpha-1}$ & $0,\ 2$\\\cline{2-7}
 \hline 
 \end{tabular}
 \par\end{centering}
 \protect\caption{ }
 \label{tab:T1(12)B_QN}
 \end{subtable}
 \begin{subtable}[b]{0.5\textwidth}
 \setlength{\tabcolsep}{0.15cm} \global\long\def\arraystretch{1.2}
 \begin{centering}
 \begin{tabular}{|c||c||c|c|c|c||c|}
 \hline 
  Model & Sol. & $X_1^F$  & $X_2^S$  & $X_3^S$  & $X_4^F$ & $\alpha$  \tabularnewline
 \hline 
 \hline 
 \multirow{3}{1.7cm}{\centering T1-3-D} 
       & I  & $1_\alpha $  & $1_\alpha $  & $1_\alpha $  & $2_{\alpha+1}$ &$0 $\\\cline{2-7}
       & II  & $2_\alpha $  & $2_\alpha $  & $2_\alpha $  & $1_{\alpha+1}$ &$\pm1 $\\\cline{2-7}
       & III  & $2_\alpha $  & $2_\alpha $  & $2_\alpha $  & $3_{\alpha+1}$ &$\pm1 $\\\cline{2-7}
       & IV  & $3_\alpha $  & $3_\alpha $  & $3_\alpha $  & $2_{\alpha+1}$ &$0 $\\\cline{2-7}
 \hline 
 \multirow{3}{1.7cm}{\centering T1-3-E} 
       & I  & $1_\alpha $  & $1_\alpha $  & $2_{\alpha-1} $  & $1_\alpha $ & $0,\ 2$ \\\cline{2-7}
       & II  & $2_\alpha $  & $2_\alpha $  & $1_{\alpha-1} $  & $2_\alpha $ & $\pm1$ \\\cline{2-7}
       & III  & $2_\alpha $  & $2_\alpha $  & $3_{\alpha-1} $  & $2_\alpha $ & $\pm1$ \\\cline{2-7}
       & IV  & $3_\alpha $  & $3_\alpha $  & $2_{\alpha-1} $  & $3_\alpha $ & $0,\ 2$ \\\cline{2-7}
 \hline 
 \end{tabular}
 \par\end{centering}
 \protect\caption{ }
 \label{tab:T13(DE)_QN}
 \end{subtable}
 \protect\caption{Possible quantum numbers for the mediators in the T1 topologies.}
 \label{tab:T1-quantum_numbers}
 \end{table}
 
\begin{table}[h]
\begin{centering}
\begin{tabular}{|c||c||c|c|c||c|}
\hline 
 Model & Sol. & $X_1^F$  & $X_2^S$  & $X_3^S$  & $\alpha$  \tabularnewline
\hline 
\hline 
\multirow{4}{1.7cm}{\centering T3-1-A} 
      & I   & $1_\alpha$  & $1_\alpha $  & $2_{\alpha-1} $ & 0,\ 2  \\\cline{2-6}
      & II  & $2_\alpha$  & $2_\alpha $  & $1_{\alpha-1} $ & $\pm1$ \\\cline{2-6}
      & III  & $2_\alpha$  & $2_\alpha $  & $3_{\alpha-1} $ & $\pm1$ \\\cline{2-6}
      & IV  & $3_\alpha$  & $3_\alpha $  & $2_{\alpha-1} $ & $0,\ 2$ \\\cline{2-6}
\hline 
\end{tabular}
\par\end{centering}
\protect\caption{Possible quantum numbers for the mediators in the T3 topology.}
\label{tab:T3-quantum_numbers}
\end{table}

\begin{table}[h]
\begin{centering}
\begin{tabular}{|c||c||c|c|c|c||c|}
\hline 
 Model & Sol. & $X_1^F$  & $X_2^F$  & $X_3^S$  & $X_4^S$ & $\alpha$  \tabularnewline
\hline 
\hline 
\multirow{2}{1.7cm}{\centering T4-3-I} 
      &IV  & $3_0$  & $2_\alpha $  & $2_\alpha $  & $2_\alpha $ & $\pm1$  \\\cline{2-7}
      & V  & $3_0$  & $3_\alpha $  & $3_\alpha $  & $3_\alpha $ &          \\\cline{2-7}
\hline 
\end{tabular}
\par\end{centering}
\protect\caption{Possible quantum numbers for the mediators in the T4 topology.}
\label{tab:T4-quantum_numbers}
\end{table}

\bibliographystyle{apsrev4-1long}
\bibliography{references}

\end{document}